\documentstyle[sprocl,epsfig,amssymb]{article}

\def\be{\begin{equation}}
\def\ee{\end{equation}}
\def\bea{\begin{eqnarray}}
\def\eea{\end{eqnarray}}
\newcommand{\kos}{\ifmmode \mathrm{K^{0}_{S}} \else K$^{0}_{\mathrm S} $ \fi}
\newcommand{\kol}{\ifmmode \mathrm{K^{0}_{L}} \else K$^{0}_{\mathrm L} $ \fi}
\newcommand{\kob}{\ifmmode {\overline{\mathrm{K}^{0}}} 
\else $\overline{\mathrm{K}^{0}} $ \fi}

\def\ifmath#1{\relax\ifmmode #1\else $#1$\fi}%

\def\rf{{\mathrm{f}}}

\def\rG{\ifmath{{\mathrm{G}}}}

\def\rL{\ifmath{{\mathrm{L}}}}

\def\rt{\ifmath{{\mathrm{t}}}}
\def\rT{\ifmath{{\mathrm{T}}}}

\def\rx{\ifmath{{\mathrm{x}}}}
\def\ry{\ifmath{{\mathrm{y}}}}

\def\vec#1{{\mbox{\bf #1}}}
\def\vs{\vskip}

\begin{document}

\title{COMPARISION OF THE PION EMISSION FUNCTION IN HADRON-HADRON
AND HEAVY-ION COLLISIONS} 

\author{ RAFAEL HAKOBYAN }
\address
{
High Energy Physics Institute (HEFIN), University of
Nijmegen/NIKHEF, NL-6525 ED Nijmegen, The Netherlands \\
Yerevan Physics Institute, AM-375036 Yerevan, Armenia  \\
E-mail: raphael@jerewan1.yerphi.am or rafael@hef.kun.nl
}

\maketitle\abstracts{
 $\quad $Combining results on single-particle distributions with those
 of the
 Bose-Einstein correlation analysis in the same experiment, the space-time
 emission function is extracted as a function of time and longitudinal
 coordinate, as well as a function of transverse coordinates.
 While the former function resembles a boomerang for both types of
 collision (only shifted in the time direction due to the larger
 freeze-out
 time in heavy-ion collisions), the latter function is Gaussian
 for heavy-ion collisions, but ring-shaped for hadron-hadron collisions.  
}

\section{\bf Introduction}
$\qquad $One of the focal points of current research  in high
energy heavy-ion and particle physics 
is the extraction of the space-time dependence of the pion emission
function of  the particle
production points.

In this paper we will consider the space-time 
distribution of
correlated pion production points in hadron-hadron and heavy-ion
collisions at the CERN SPS.
It is interesting to note, that the hydrodynamical models not only
hold for heavy-ion collisions,[1--4] but also can be used for
meson-proton collisions
[5--8] for which the formation of
hadronic matter with macroscopic properties is less evident.  
It is important to emphasize, that correlation measurements do
not contain the complete information on the geometrical and dynamical
parameters characterizing the evolution of hadronic matter. More
profound information can be provided by a combined analysis of data on 
two-particle correlations and single-particle inclusive 
spectra.[7--16] 

Here, the results of an analysis are presented on
($\pi^+$/K$^+$)p[5--8] and Pb+Pb interactions[2--4,17--19,25--28] at the
CERN SPS 
in the framework of the hydrodynamical 
model for three-dimensionally expanding, cylindrically symmetric, finite
systems (BL-H model).[11] For both types of collision, the space-time
emission function is extracted as a function of time and longitudinal
coordinate, as well as a function of transverse coordinates. 
\vfill\eject

\section{\bf
Combination of Single-Particle Spectra and Two-Particle
Correlations
}

$\qquad $The Bertsch-Pratt (BP) parametrization [21--24] has been
used to describe the 
two-particle correlation function. At mid-rapidity, $y=y_0$, 
and
in the Longitudinal Center of Mass System (LCMS), where the longitudinal
momentum sum is zero, the effective BP radii can be approximately
expressed from the BL-H parametrization as [11,14,20,21]
\be
R_\rL^2=\tau_\rf^2\Delta \eta_*^2 ; \qquad 
R_0^2=R_*^2+\beta_\rt^2\Delta \tau_*^2 ;  \qquad
R_s^2=R_*^2
\ee        
with
\be
\frac{1}{\Delta \eta_*^2}=\frac{1}{\Delta \eta^2}+\frac{M_\rt}{T_0} ;
\qquad
R_*^2=\frac{R_\rG^2}{1+\frac{M_\rt}{T_0}(\langle u_\rt\rangle^2+\langle 
\frac{\Delta T}{T}\rangle)}, \\
\ee
where $M_\rt=0.5(m_{\rT,1}+m_{\rT,2})$ and 
the parameters $\Delta \eta^2,T_0,\langle
u_\rt\rangle$ and $\langle \frac{\Delta T}{T}\rangle $  are extracted 
from the invariant spectra;[11]
$R_\rG$ is related to the transverse geometrical r.m.s. radius of the 
source as $R_\rG$(r.m.s.)=$\sqrt{2} R_\rG$;
$\tau_\rf$ is the mean freeze-out (hadronization) time;
$\Delta \tau_*$ is related to the duration time $\Delta \tau$ of hadron 
emission and to the temporal inhomogeneity of local temperature, as the
relation $\Delta \tau_* \geq \Delta \tau $ holds;
the variable $\beta_\rt$ is the transverse velocity of the pion pair. 

\vs 4mm
\begin{center}
\begin{tabular}{|p{1.5cm}|p{1.5cm}p{1.5cm}p{1.5cm}p{1.5cm}|p{1.7cm}|}
\hline
Param. &  NA49 & NA44 & WA98 & Averaged & NA22 \\ 
\hline
$T_0$ [MeV]  & 134$\pm$3 & 145$\pm$3 & 139$\pm$5 & 139$\pm$6 
& 140$\pm$3 \\
$\langle u_t\rangle$ & 0.61$\pm$0.05 & 0.57$\pm$0.12 & 0.50$\pm$0.09 
& 0.55$\pm$0.06 & 0.20$\pm$0.07 \\
$R_G$ [fm] & 7.3$\pm$0.3 & 6.9$\pm$1.1 & 6.9$\pm$0.4 & 7.1$\pm$0.2 
& 1.2$\pm$0.2 \\
$\tau_0$ [fm/c] & 6.1$\pm$0.2 & 6.1$\pm$0.9 & 5.2$\pm$0.3 & 5.9$\pm$0.6
& 1.4$\pm$0.1 \\
$\Delta\tau$ [fm/c] & 2.8$\pm$0.4 & 0.01$\pm$2.2 & 2.0$\pm$1.9 &
1.6$\pm$1.5 & 1.3$\pm$0.3 \\
$\Delta\eta$ & 2.1$\pm$0.2 & 2.4$\pm$1.6 & 1.7$\pm$0.1 & 2.1$\pm$0.4 
& 1.36$\pm$0.02 \\
$\langle {\Delta T \over T}\rangle$ & 0.07$\pm$0.02 & 0.08$\pm$0.08 
& 0.01$\pm$0.02 & 0.06$\pm$0.05 & 0.71$\pm$0.14 \\
$y_0$ & 0 (fixed) & 0 (fixed) & 0 (fixed) & 0 & 0.082$\pm$0.006 \\
\hline
$\chi^2/NDF$ & 163/98 & 63/71 & 115/108 & & 642/683 \\
\hline
\end{tabular}
\end{center}
\vs 2mm
{\small\baselineskip=10pt\noindent
 TABLE 1. $\,\,\,$ Fit paramaters of the Buda-Lund hydro (BL-H) model
in a combined (simultaneous for heavy-ion collisions) analysis of
NA22, NA49, NA44, WA98 spectra and correlation data 
(the 5th column corresponds to the weighted average of the previous
three columns
and the error is composed by the statistical and systematic ones, while
in the 6th column only statistical errors are presented).\par}
\vs 4mm

  The fit parameters of the combined analysis of the single-particle
spectra
and the two-particle Bose-Einstein correlation functions in
($\pi^+$/K$^+$)p
interac- tions,[5--8] 
as well as the fit parameters of the simultaneous 
analysis
of the BL-H model to particle correlations and spectra in Pb+Pb 
collisions[2--4,17--19,25--28,30]
are presented in Table~1.

\section{\bf Emission Function Expressed as a Function of Time and
Longitudinal Coordinates}

$\qquad $The momentum-integrated emission function along the $z$-axis,
i.e., at
${\vec r}_\rt=(r_\rx,r_\ry)=(0,0)$ is given by

\be
 S(t,z) \propto \exp\left(-  {(\tau - \tau_0)^2\over 2 \Delta \tau^2}
\right) \exp\left( - {(\eta - y_0)^2   \over 2 \Delta \eta^2} \right), 
\ee
where the parameters are taken from Table~1.

The reconstruction of the space-time distribution of hadron emission
points
is shown in Fig.~1, where the upper-left and lower-left diagrams 
correspond to
the reconstructed $S(t,z)$ emission function in arbitrary units, as a
function of the cms time variable $t$ and the cms longitudinal coordinate
$z\equiv r_z$, for ($\pi^+$/K$^+$)p and Pb+Pb interactions,
respectively.
Note, the coordinates ($t,z$), are expressed with the help of the
longitudinal proper-time $\tau $ and space-time rapidity $\eta$ as 
$(\tau \cosh \eta, \tau \sinh \eta )$. 

In both types of reactions, the space-time structure 
resembles a boomerang, i.e., particle 
production takes place close to $\mid z\mid \approx t$,
with gradually decreasing probability for ever larger values of space-time
rapidity.
We see a characteristic long tail of particle emission on both sides of
the light cone, giving at least 40 fm longitudinal extension in
$z$ and 20 fm/$c$ duration of particle production in
$t$ for h-h collisions, while for A-A collisions they are
longer than  150 fm
and 80 fm/$c$, respectively. We see a sharper freeze-out
hypersurface in
heavy-ion than in hadron-proton collisions, indicating that in 
hadron-proton interactions the emission process occurs during almost all
the
hydrodynamical evolution (the duration time of pion emission , $\Delta
\tau = 1.3 \pm 0.3 $, is close to the mean freeze-out time, 
$\tau_\rf =1.4\pm 0.1$
fm/$c$), while in A-A collisions a large mean freeze-out time,
$\tau_\rf =5.9\pm 0.6$ fm/$c$, is
found with a relatively short duration of emission,
$\Delta \tau = 1.6 \pm 1.5$ fm/$c$.
Note, that the temporal
cooling in Pb+Pb collisions seems to be stronger than in h+p collisions,
which
can be explained by a faster three-dimensional expansion in the former
case, as compared to the essentially one-dimensional expansion in the case
of h+p interactions.  

%\begin{figure}
\begin{center}
\epsfig{figure=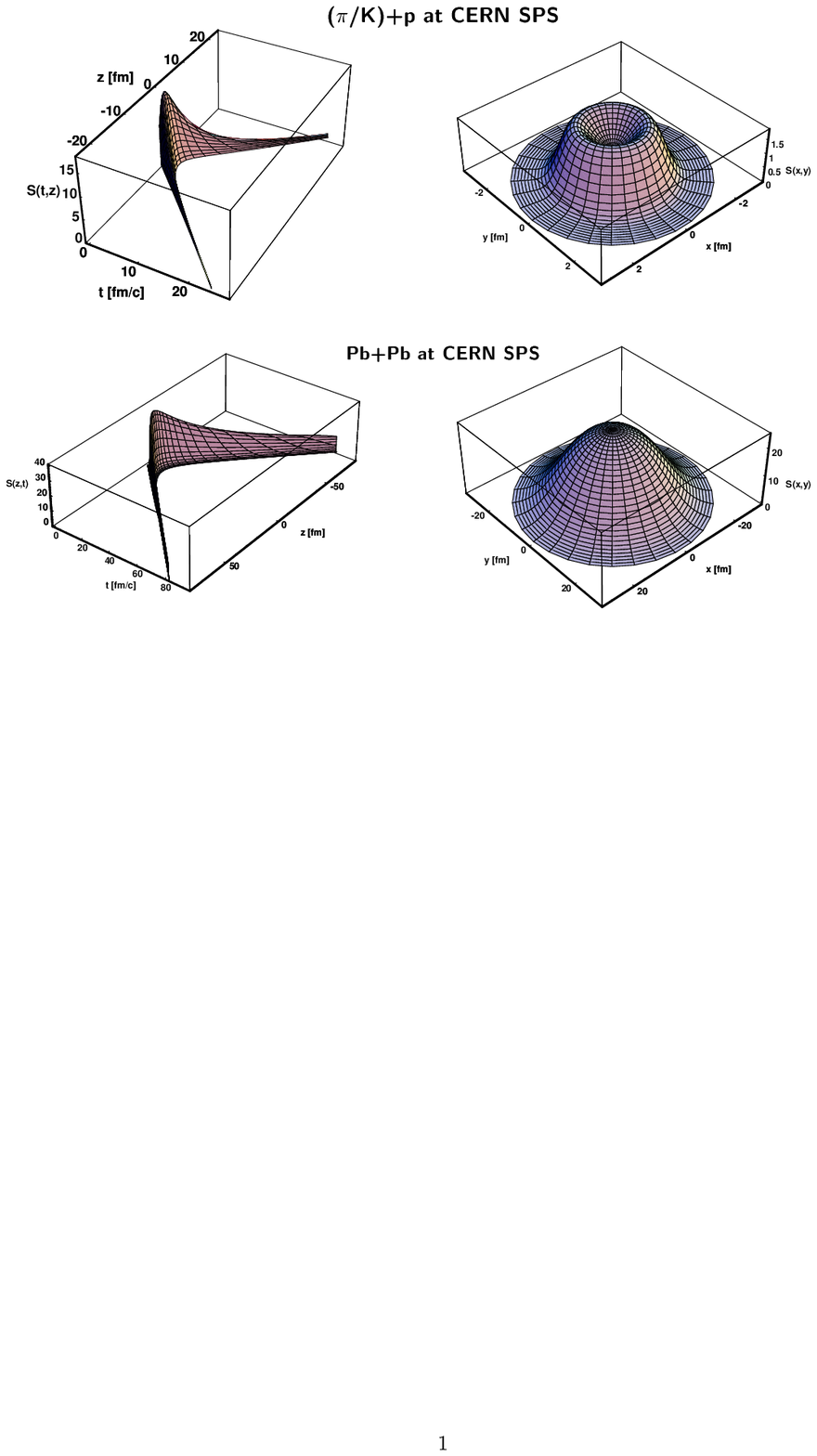,width=12cm}
\end{center}
\vs -1mm
{\small\baselineskip=10pt\noindent
Figure 1: 
The reconstructed $S(t,z)$ emission function in arbitrary vertical units,
as
a function of time $t$ and longitudinal coordinate $z$ (left diagrams),
as well as the reconstructed $S(x,y)$ emission function in arbitrary
vertical units, as a function of the transverse coordinates
$x\equiv r_\rx$ 
%$x\equiv r_x$ 
and $y\equiv r_\ry$
%and $y\equiv r_y$
(right), for h-h (upper) and A-A (lower) collisions, respectively. 
\par}
\vs 4mm
%\end{figure}

\section{\bf Emission Function Expressed as a Function of Transverse
Coordinates}

$\qquad $In the transverse direction, only the rms width of the source
can
be
directly inferred from the BP radii. However, the additional information
on the values of $\langle u_\rt\rangle $ and on 
the values of $\langle \frac{\Delta T}{T}\rangle $ 
from the analysis of the transverse momentum distribution 
can be used to reconstruct the details of the transverse
density profile. An exact, non-relativistic hydro solution was found in
ref. [29], in terms of the parameters $\langle u_\rt\rangle $ and 
$\langle \frac{\Delta T}{T}\rangle $ using an ideal gas equation of
state. In this hydro solution
\be
\langle \frac{\Delta T}{T}\rangle \gtrless \frac{m\langle 
u_\rt\rangle^2}{T_0}  ,
\ee
where the $<$ sign corresponds to a self-similar expanding fire-ball, 
while 
%\be
%\langle \frac{\Delta T}{T}\rangle > \frac{m\langle u_\rt\rangle^2}{T_0},
%\ee
the opposite
$>$ sign corresponds to a self-similar expanding ring of fire. 
Assuming the validity of this non-relativistic solution in the transverse
direction, one can reconstruct the detailed
shape of the transverse density profile. 
The result looks like a ring of fire in the $(r_x, r_y)$ plane in h+p
interactions (see upper-right figure of Fig.~1), while in A+A collisions
it has a Gausian distribution (see lower-right figure of Fig.~1). 

So, we
conclude that the pion emission function $S(r_x,r_y)$ in h+p collisions
corresponds to the formation of a ring of fire in the transverse plane.
This is due to the rather small transverse flow and the
sudden drop
of the temperature in the transverse direction, which leads to large
pressure gradients in the center and small pressure gradients and a
density augmentation at the expanding radius of the fire-ring. 
This transverse
distribution, together with the scaling longitudinal expansion, creates an
elongated, tube-like source in three dimensions, with the density of
particle production being maximal on the surface of the tube. 

The pion
emission function $S(r_x,r_y)$ in A+A collisions corresponds to the radial
expansion, which is a well established phenomenon in heavy-ion collisions
from low-energy to high-energy reactions.
%[31--36] 
This transverse
distribution, together with the scaling longitudinal expansion, creates 
a cylindrically symmetric, large and transversally
homogeneous fireball, expanding three-dimensionally with a large mean
radial component $\langle u_\rt\rangle $ of hydrodynamical four-velocity.

Because of this large difference observed for those two types of
collision, analysis of the emission function in e$^+$e$^-$ collisions is
important and has been started. 

\section{\bf Acknowledgments}
\vskip 3 mm
$\qquad $I am thankful to T. Cs\"org\H{o}, W. Kittel and H. Gulkanyan
for constructive and useful discussions. 
I am grateful to members of the NA22, NA44 and NA49
Collaboration.
I would like to thank all organizers of the 30-th International
Symposium on Multiparticle Dynamics (ISMD 2000)
for their hospitality and assistance.


\section*{References}
\begin{thebibliography}{99}
\itemsep 0pt
\parsep 0pt
\bibitem{a} T. Alber {\em et al} (NA35 Coll.), {\em Z. Phys.} C {\bf 66}, 77 
(1995).
\bibitem{b} A. Franz {\em et al} (NA44 Coll.), {\em Nucl.Phys.} A {\bf 610},
240c (1996). 
\bibitem{c} K. Kadija (NA49 Coll.), {\em Nucl. Phys.} A {\bf 610},
248c (1996). 
\bibitem{d} L. Rosselet {\em et al} (WA98 Coll.), {\em Nucl. Phys.} A 
{\bf 610}, 256c (1996). 
\bibitem{e} N.M. Agababyan {\em et al} (NA22 Coll.), {\em Z. Phys.} C
{\bf 71}, 405 (1996). 
\bibitem{f} H. Gulkanyan, R. Hakobyan, W. Kittel in
{\em Proc. of the 7th Int. 
Workshop on Multiparticle Production}, Nijmegen 1996,
ed. R.C. Hwa, W. Kittel, W.J. Metzger, D.J. Schotanus (World Scientific,
Singapore, 1997) p.26.
\bibitem{h} N.M. Agababyan {\em et al} (NA22 Coll.), {\em Phys. Lett.} B
{\bf 422}, 359 (1998).
\bibitem{i} R. Hakobyan in {\em Proc. Correlations and Fluctuations$\prime$98, 
M\'atrah\'aza 1998}, ed. T.Cs\"org\H{o}, S.Hegyi, G.Jancs\'o
and R.C.Hwa,
(World Scientific, Singapore, 1999), p.128.
\bibitem{j} S. Chapman, J.R. Nix, U. Heinz, {\em Phys. Rev.} C {\bf 52},
2694 (1995). 
\bibitem{k} S.V. Akkelin, Yu.M. Sinyukov, {\em Z. Phys.} C {\bf 72},
{\bf 501} (1996). 
\bibitem{l} T. Cs\"org\H{o}, B. L\"orstad, {\bf Phys. Rev.} C {\bf 54}, 
1390 (1996). 
\bibitem{m} S.V. Akkelin, Yu.M. Sinyukov, {\em Phys. Lett.} B {\bf 356}, 
525 (1995). 
\bibitem{n} T. Cs\"org\H{o}, B. L\"orstad and J. Zim\'anyi, {\em Phys. Lett.}
B {\bf 338}, 134 (1994).
\bibitem{o} T. Cs\"org\H{o}, {\em Phys. Lett.} B {\bf 347}, 354 (1995). 
\bibitem{p} T. Cs\"org\H{o}, B. L\"orstad, {\em Nucl. Phys.} A {\bf 590}, 
465 (1995). 
\bibitem{q} T. Cs\"org\H{o} and B. L\"orstad, hep-ph/9511404, in
{\em Proc. XXV-th
Int. Conf. Multiparticle Dynamics}, Stara Lesna, Slovakia, 1995, 
ed. D.Bruncko et al., (World
Scientific, Singapore, 1996), p.661
\bibitem{r} A. Ster, T. Cs\"org\H{o} and B. L\"orstad, hep-ph/9809571,
in {\em Proc. Correlations and Fluctuations$\prime$98, M\'atrah\'aza}, Hungary,
June 1998, ed. T.Cs\"org\H{o}, S.Hegyi, G.Jancs\'o and R.C.Hwa,
(World Scientific, Singapore, 1999), p. 108.
\bibitem{s} A. Ster, T. Cs\"org\H{o} and J. Beier, {\em Heavy
Ion Phys.} {\bf 10}, 85 (1999).
\bibitem{t} A. Ster, T. Cs\"org\H{o} and B. L\"orstad,
{\em Nucl. Phys.} A (2000) in press
\bibitem{v} T. Cs\"org\H{o}, S. Nickerson, D. Kiang in
{\em Proc. 7th Int. 
Workshop on Multiparticle Production}, Nijmegen 1996,
ed. R.C. Hwa, W. Kittel, W.J. Metzger, D.J. Schotanus (World Scientific,
Singapore, 1997) p.50.
\bibitem{u} G. Bertsch, M. Gong, M. Tohyana, {\em Phys. Rev.} C {\bf 37}, 
1896 (1988).  
\bibitem{w} G.F. Bertsch, {\em Nucl. Phys.} A {\bf 498}, 173c (1989).  
\bibitem{x} G.F. Bertsch and G.E. Brown, {\em Phys. Rev.} C {\bf 40},
1830 (1989).  
\bibitem{y} S. Pratt, {\em Phys. Rev.} D {\bf 33}, 1314(1986).  
\bibitem{z} I.G. Bearden {\em et al} (NA44 Coll.), {\em Phys. Rev. Lett.}
{\bf 78}, 2080 (1997).
\bibitem{ } H. Beker {\em et al} (NA44 Coll.), {\em Phys. Rev. Lett.} {\bf 74},
3340 (1995) and {\em Nucl. Phys.} A {\bf 566}, 115c (1993).
\bibitem{ } H. Appelsh$\ddot{a}$user {\em et al} (NA49 Coll.), 
{\em Phys. Lett.} B {\bf 467}, 21 (1999).
\bibitem{ } H. Appelsh$\ddot{a}$user {\em et al} (NA49 Coll.), 
{\em Phys. Rev. Lett.} {\bf 82}, 2471 (1999).
\bibitem{ } T. Cs\"org\H{o}. nucl-th/9809011.
\bibitem{ } T. Cs\"org\H{o} in {\em Proc. Particle Production Spanning MeV
and TeV Energies}, ed. W.Kittel, P.J. Mulders and O. Scholten, 
(NATO Science Series, 2000), p.203.
%\bibitem{ } W. Bauer, C.-K. Gelbke and S. Pratt, {\em Annu. Rev. Nucl. 
%Part.
%Sci.} {\bf 42}, 77 (1992).
%\bibitem{ } D. Ardouin, {\em Int. J. Mod. Phys.} E {\bf 6}, 391 (1997).
%\bibitem{ } N. Herrmann {\em et al} (FOPI Coll.), {\em Nucl. Phys.} A
%{\bf 610}, 49 (1996).
%\bibitem{ } B. Hong {\em et al} (FOPI Coll.), {\em Phys. Rev.} C {\bf 
%57},
%244 (1998).
%\bibitem{ } P. Crochet {\em et al} (FOPI Coll.), {\em Nucl. Phys.} A
%{\bf 624}, 755 (1997).
%\bibitem{ } R. Kotte {\em et al} (FOPI Coll.), {\em Eur. Phys. J.} A {\bf 
%6}, 
%185 (1999).
\end{thebibliography}
\end{document}

%%%%%%%%%%%%%%%%%%%%%%
% End of sprocl.tex  %
%%%%%%%%%%%%%%%%%%%%%%